# Longitudinal magnon, inversion breaking and magnetic instabilities in the pseudo-Kagome francisites $Cu_3Bi(SeO_3)_2O_2X$ with $X$=Br, Cl

V. Gnezdilov[1,2], Yu. Pashkevich[3], V. Kurnosov[2], P. Lemmens[1],
E. Kuznetsova[4], P. Berdonosov[4], V. Dolgikh[4], K. Zakharov[4], and A. Vasiliev[4,5,6]

[1] Inst. Cond. Matter Physics, Univ. Technology Carolo Wilhelmina, D-38106 Braunschweig, Germany

[2] B. Verkin Institute for Low Temp. Physics and Engineering NASU, Kharkov, Ukraine

[3] O. Galkin Donetsk Institute for Physics and Engineering NASU, Kiev, Ukraine

[4] MV Lomonosov Moscow State University, 119991 Moscow, Russia

[5] Ural Federal University, 620002 Ekaterinburg, Russia

[6] National University of Science and Technology "MISiS", 119049 Moscow, Russia

We performed Raman studies and a dielectric characterization of the pseudo-kagome $Cu_3Bi(SeO_3)_2O_2X$ ($X$ = Cl, Br). These compounds share competing nearest-neighbour ferromagnetic exchange and frustrating next-nearest-neighbour antiferromagnetic exchange as well as highly noncollinear magnetic ground state. However, at low temperature they differ with respect to the existence of inversion symmetry. For both compounds there exists a pronounced interplay of polar phonon modes with quantum magnetic fluctuations. A novel Raman mode appears for temperatures below the Neel temperature with a Fano lineshape and an enormous intensity that exceeds most of the phonon lines. We discuss a possible contribution of longitudinal magnons to this signal. In contrast, one magnon scattering based on linear transvers magnons is excluded based on a symmetry analysis of spin wave representations and Raman tensor calculations. There exists evidence that in these pseudo-kagome compounds magnetic quantum fluctuations carry an electric dipole moment. Our data as well as a comparison with previous far-infrared spectra allow us to conclude that $Cu_3Bi(SeO_3)_2O_2Cl$ changes its symmetry most likely from *Pmmn* to *P2₁mn* with a second order structural phase transition at $T^*$=120 K and becomes multiferroic. $Cu_3Bi(SeO_3)_2O_2Br$ represents an interesting counter part as it does not show this instability and stays inversion symmetric down to lowest temperatures, investigated.

PACS:

## I. Introduction

The $S = 1/2$ kagome-lattice Heisenberg antiferromagnets have a conceptual importance in the study of frustrated magnetism, because they are based on triangles of competing exchange paths as basic units, related to strong quantum fluctuations. In these systems the ground state is a superposition of highly degenerate and entangled quantum states that fluctuate down to lowest temperatures, i.e., a spin liquid. Despite this interest in spin liquids in solid states there are very few two-dimensional magnets which fail to order at low temperatures. On the other hand the effect of lattice distortions and anisotropies on the magnetic exchange topology and the fluctuation spectrum in the kagome lattice is an interesting problem in itself. While most of the interest in the ideal kagome lattice is associated with antiferromagnetic interactions, it was shown that frustration due to competing exchange can have a dramatic impact even if the dominant exchange couplings are ferromagnetic. Exactly this situation is realized in a series of layered cuprates $Cu_3Bi(SeO_3)_2O_2X$ ($X$ = Br, Cl).[1]

$Cu_3Bi(SeO_3)_2O_2X$ crystallizes in an orthorhombic symmetry with the space group *Pmmn*.[2] The system is composed of $Cu^{2+}$ ions on slightly buckled kagome layers with weak interlayer coupling.[3] The structure is characterized by two symmetry inequivalent copper sites, Cu1 (4c) and Cu2 (2a). These copper sites form corner shared [$CuO_4$] units that are additionally linked by $Se^{4+}$ and $Bi^{3+}$ ions. The $Cl^-$ and $Br^-$ ions sandwich the planes formed by the copper hexagons. A more detailed descriptions of the crystal structure and pictures can be found in Refs. (1, 2, 3, 4).

Earlier experimental investigations revealed: (i) almost identical magnetic properties for the bromide and chloride compounds with the magnetic ordering temperatures 27.4 and 24 K, respectively,[3,4,5,6] (ii) two magnetic modes at low temperatures at 1.23 meV (9.92 cm$^{-1}$) and 1.28 meV (10.32 cm$^{-1}$) in $X$ = Br[5] and two modes at 1.18 meV (9.5 cm$^{-1}$) and 4.1 meV (33.1 cm$^{-1}$) in $X$ = Cl[6], and (iii) the existence of a structural phase transition with symmetry lowering at $T \approx 115$ K in $X$ = Cl.[6]

The magnetic structure has already been discussed in Refs. 4 and 5 for the $X$ = Br compound. The proposed magnetic structure consists of the Cu1 moments which are aligned parallel to the *c* axis with additional alternating components along the *b* axis and are thus canted nearly ±50° from *c* towards *b*. The Cu2 moments are strictly parallel to the *c* axis. Units of kagome magnetic structure are corner shared triangles; each is formed by two Cu1 and one Cu2. The nearest-neighbour (NN) interactions between all copper ions are FM. However, there is sizable frustrating AFM Cu1-Cu1 interaction from different triangles disposed along the *b*-axis. This next-nearest-neighbour (NNN) interaction is realized through Cu1-O-Bi-O-Cu1 paths, meanwhile the presence of Bi ions provides that AFM NNN-interaction becomes even larger than FM NN one.[1,4] The competition between FM and AFM interactions is responsible for the huge canting of Cu1 moments from the *c*-axis, which, therefore, does not have a relativistic origin. The coupling between the kagome layers is AFM and due to this magnetic unit cell is equal to doubled crystallographic one along *c*-axis. Another cause for spin canting is Dzyaloshinsky-Moriya (DM) interaction which was found to be finite on all bonds. According to the density-functional band-structure calculations, the DM vector on the Cu1-Cu2 bonds is the most important ingredient and must be accounted for to explain all properties of the kagome francisites qualitatively.[1]

The improvement of microscopic magnetic models for francisites with kagome-like structure requires further experimental investigations. Several quantitative features point to a more complex nature of these materials. Especially it is needed to clarify the crystallographic symmetry and the nature of the putative magnetic or structural transformations in the temperature region of 100–150 K. According to Ref. 3, the slope of the inverse magnetic susceptibility changes at $T$ = 150 K with two sets of a Curie–Weiss parameters above and below this temperature.

Our approach of investigating the excitations in the kagome-like lattice is using Raman scattering. This is a unique method in the sense that it simultaneously probes different degrees of freedom and their interaction. In particular, Raman scattering provides a way to distinguish possible ground states.[7,8] Recently, this technique has been successfully applied to characterize the excitation spectrum of the spin liquid state in a Heisenberg $S$ = 1/2 antiferromagnet on a perfect kagome lattice – herbertsmithite $ZnCu_3(OH)_6Cl_2$.[9] Investigations of two compounds volborthite ($Cu_3V_2O_7(OH)_2 \cdot 2H_2O$) and vesignieite ($BaCu_3V_2O_8(OH)^2$) with slightly distorted, kagome planes demonstrate that even small modifications of the crystal structure may have a huge effect on its properties.[10]

Here, we report on a detailed study of the temperature dependence of the vibrational and magnetic excitations in $Cu_3Bi(SeO_3)_2O_2X$ ($X$ = Cl, Br). Data analysis allows us to affirm that the second order phase transition at $T^*$ = 120 K in $Cu_3Bi(SeO_3)_2O_2Cl$ occurs with a loss of inversion symmetry. Raman data allow us to conclude the non-centrosymmetric orthorhombic space group $P2_1mn$ as the low-temperature phase of this compound. A low-frequency, highly intensive scattering signal of magnetic origin is observed at low temperatures in both Cl- and Br-containing compounds. The origin of the magnetic signal is discussed from the point of view of Fano resonance and scattering on longitudinal-magnons.[11,12,13] The symmetry analysis supports both mechanisms. In addition, the anomalies observed in the magnetic specific heat $C(T)$ and the permittivity $\varepsilon(T)$ confirm our conclusions.

## II. EXPERIMENTAL DETAILS

Single crystals of $Cu_3Bi(SeO_3)_2O_2Cl$ were obtained by using chemical transport in temperature gradients of 480/400°C, as reported in Ref. 6. The selenium dioxide was obtained from $H_2SeO_3$ (98%) by dehydration under vacuum and gentle heating by hot water (about 80-70°C). The intermediate product has been additionally purified by sublimation in flow of dry air and $NO_2$. The mixture of ultra pure CuO 0.2951 g (3.7 mmol), reagent grade $CuCl_2·2H_2O$ 0.0579 g (0.34 mmol), reagent grade BiOCl 0.3525 g (1.35 mmol) and anhydrous $SeO_2$ 0.297 g (2.68 mmol) was loaded under argon atmosphere into the quartz tube of 17 cm length and 1.8 cm diameter. The tube has been sealed under vacuum about 0.03-0.04 tor. The composition of mixture was chosen according to the supposed reaction 2.75 CuO + BiOCl + $2SeO_2$ + 0.25 $CuCl_2·2H_2O$ = $Cu_3Bi(SeO_3)_2O_2Cl$ + 0.5HCl + $0.25H_2O$.
The tube was placed into horizontal two zone furnace and both ends of the tube were heated up to 300°C for 6 hrs. After the 20 hrs of exposition at 300°C the temperatures were raised to the 400°C. After that, the evaporation zone with starting mixture was heated up to 480°C for 5 hrs. The transportation time was 18 days. As the result, the green well-shaped crystals with dimension of 3-5 mm were synthesized in the cold tube zone.

The $Cu_3Bi(SeO_3)_2O_2Br$ crystals were obtained from a stoichiometric mixture of ultra pure CuO 0.3223 g (4 mmol), reagent grade BiOBr 0.4123 g (1.35 mmol) and $SeO_2$ 0.304 g (2.7 mmol). The mixture was loaded into a quartz tube of 17 cm length and 1.8 cm diameter and a small drop of dried bromine (99,9+) was added. The transport crystal growth was performed in a horizontal two zone furnace. The tube was heated up to 400°C for 6 hrs and after exposition at 400°C the temperature of the hot zone with the original mixture was raised up to 480°C for 6 hrs. The transportation time was 30 days. The crystals of $Cu_3Bi(SeO_3)_2O_2Br$ were found in the cold zone of the transport tube. It should be noted that the previously reported temperature gradient of 550-500 °C in Refs. 4 and 5 for $Cu_3Bi(SeO_3)_2O_2Br$ leads to a molten product in the condensation zone.

The green crystals from both compositions were crunched into fine powder and tested by powder XRD on a DRON-3M setup employing Co Kα radiation. The patterns were indexed in the *Pmmn* (#59) orthorhombic space group with cell constants for $Cu_3Bi(SeO_3)_2O_2Cl$ $a$ = 6.341(5), $b$ = 9.641(10) and $c$ = 7.222(6) Å, cell volume 441.5(9) Å$^3$; for $Cu_3Bi(SeO_3)_2O_2Br$ $a$ = 6.377(5), $b$ = 9.681(11), and $c$ = 7.268(4) Å, cell volume 448.7(9) Å$^3$. These values are in good agreement with the previously reported ones.[2,3]

The temperature dependences of magnetic susceptibility in $Cu_3Bi(SeO_3)_2O_2X$ ($X$ = Cl, Br) single crystals have been investigated in Refs. 4, 6. Both compounds order antiferromagnetically at comparable Neel temperatures $T_N$ of about 27 K. Below $T_N$, both compounds exhibit pronounced magnetic anisotropy and metamagnetic transition at $B \parallel c$ axis.

The specific heat measurements (see Fig. 1 (a)) confirm the formation of long range order in $Cu_3Bi(SeO_3)_2O_2X$ ($X$ = Cl, Br) with a Neel temperature in Br compound being slightly higher (~ 0.5 K) than that in the Cl compound. At $T_N$, both systems evidence sharp $\lambda$-type anomalies expected for a second order phase transition. At lowest temperatures, far below the phase transition the $C/T$ vs. $T^2$ dependences are practically linear in both compounds, as shown in the Inset of Fig. 1(a). Basically, the slope of these dependencies defines the Debye temperature. However, in the case of an antiferromagnetically ordered compound the phonon and magnon contributions follow the $T^3$ law being therefore additive. So, only the lower limit of Debye temperature $\Theta_D = 185 \pm 15$ K can be estimated for $Cu_3Bi(SeO_3)_2O_2X$ ($X$ = Cl, Br). An additional anomaly at $T^* \sim 119$ K exists in the temperature dependence of the specific heat in the Cl compound. This anomaly is not observed in the magnetization and can be ascribed to the structural phase transition, as found in infrared (IR) spectroscopy[6] and birefringence measurements.[3]

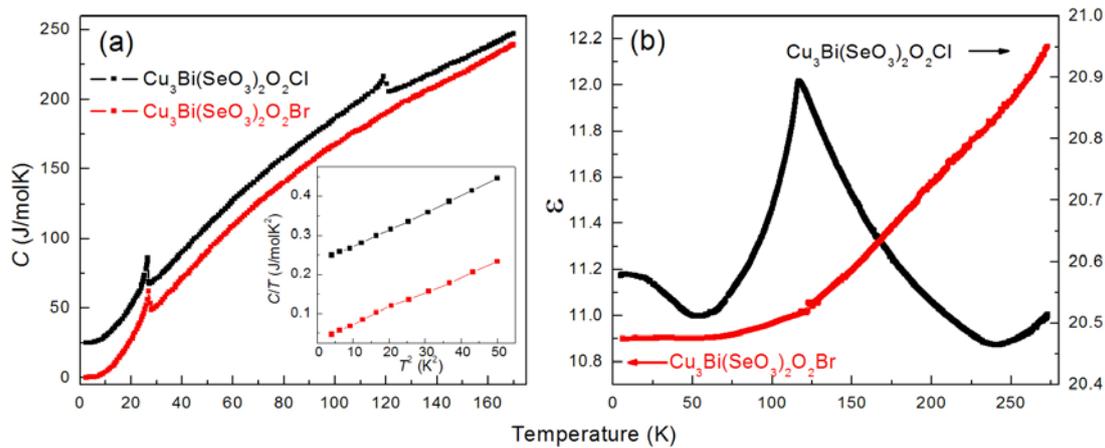

Fig. 1. (a) Specific heat and (b) dielectric permittivity of $Cu_3Bi(SeO_3)_2O_2X$ ($X$ = Cl, Br) single crystals as function of temperature. The inset in (a) indicates a $C/T$ vs. $T^2$ dependence for $T < 7$ K. The curves in (a) main panel are shifted with respect to each other by 25 J/molK and for 0.25 J/molK$^2$ in the inset of (a).

The measurements of permittivity in $Cu_3Bi(SeO_3)_2O_2X$ further confirm the presence of a structural phase transition in $Cu_3Bi(SeO_3)_2O_2Cl$. Fig. 1 (b) shows $\varepsilon(T)$ for both title compounds with a $\lambda$-shaped peak for $Cu_3Bi(SeO_3)_2O_2Cl$. This peak has a maximum at $T_{max} = 117$ K which nearly coincides with the anomaly at $T^*$ in $C(T)$.

Freshly cleaved plate-like samples with size of about 3 × 3 × 1 mm$^3$ were used for the Raman scattering investigations. Experiments were performed in quasi-backscattering geometry, using a $\lambda$ = 532-nm solid-state laser. The laser power was set to $P < 2$ mW with a spot diameter of approximately $d$=100 $\mu$m to avoid heating effects. Raman spectra were measured in both parallel ($XX$, $X'X'$) and crossed ($XY$, $X'Y'$) polarizations, with light polarized in the basal ($ab$) plane. All measurements were carried out in a closed-cycle cryostat (Oxford/Cryomech Optistat) in the temperature range from 8 to 295 K. The spectra were collected via a triple spectrometer (Dilor-XY-500) by a liquid nitrogen cooled CCD (Horiba Jobin Yvon, Spectrum One CCD-3000V).

## III. EXPERIMENTAL RESULTS and DISCUSSION

For temperatures $T > T_N$, the factor group analysis for the orthorhombic space group *Pmmn* yields 78 active and 9 silent optical modes:

$$\Gamma^{optical} = 12A_g^R + 6B_{1g}^R + 9B_{2g}^R + 12B_{3g}^R + 14B_{1u}^{IR} + 14B_{2u}^{IR} + 11B_{3u}^{IR} + 9A_{1u},$$

where indexes *R* and *IR* denote the Raman and infrared active modes, respectively. The corresponding Raman tensors are given by:

$$A_g = \begin{pmatrix} a & 0 & 0 \\ 0 & b & 0 \\ 0 & 0 & c \end{pmatrix}, B_{1g} = \begin{pmatrix} 0 & d & 0 \\ d & 0 & 0 \\ 0 & 0 & 0 \end{pmatrix}, B_{2g} = \begin{pmatrix} 0 & 0 & e \\ 0 & 0 & 0 \\ e & 0 & 0 \end{pmatrix}, B_{3g} = \begin{pmatrix} 0 & 0 & 0 \\ 0 & 0 & f \\ 0 & f & 0 \end{pmatrix}.$$

Fig. 2 shows spectra taken at room temperature (RT, $T = 295$ K) from the *ab* plane. We identify 18 and 17 $A_g + B_{1g}$ phonon modes for the $Cu_3Bi(SeO_3)_2O_2Br$ and $Cu_3Bi(SeO_3)_2O_2Cl$, respectively. This is in excellent agreement with the expected 18 Raman-active phonons.

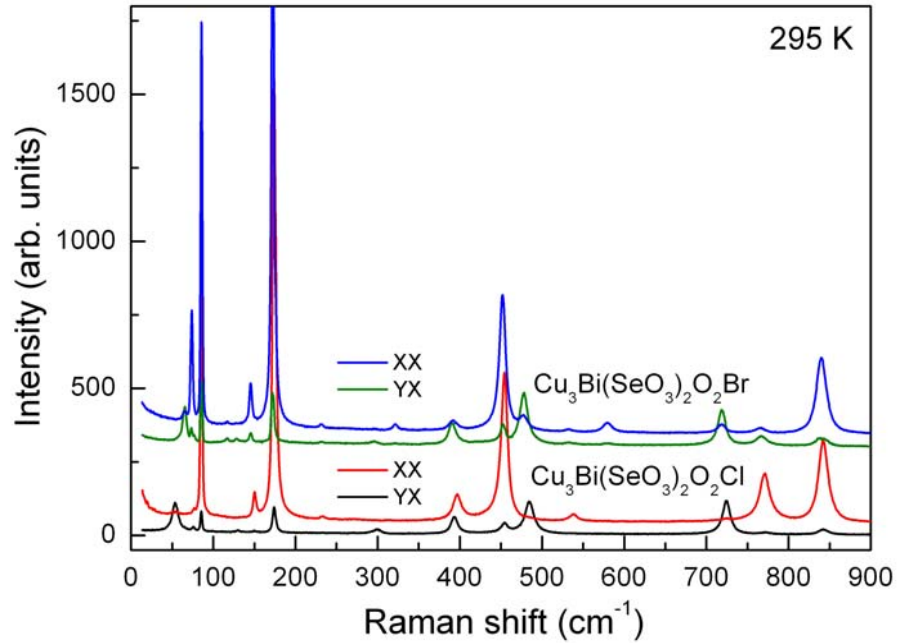

Fig. 2. Raman spectra of single crystal $Cu_3Bi(SeO_3)_2O_2Br$ and $Cu_3Bi(SeO_3)_2O_2Cl$ collected at room temperature.

The narrow line shape of the phonon lines gives evidence for a high sample quality. Furthermore, the spectra are well polarized and we can easily determine the symmetry of each phonon modes. The most intensive lines in the XX spectra with frequencies at around 173 and 453 cm$^{-1}$ have $A_g$ scattering symmetry. The lowest frequency lines in the spectra have $B_{1g}$ symmetry.

Most experiments were carried out using $X'X'$ and $Y'X'$ scattering geometries with the sample rotated by some angle around to the $c$ axis. The appropriate Raman tensors for the scattering from $ab$ plane are:

$$A_g = \begin{pmatrix} a' & d' & 0 \\ d' & b' & 0 \\ 0 & 0 & c \end{pmatrix}, B_{1g} = \begin{pmatrix} e' & f' & 0 \\ f' & -e' & 0 \\ 0 & 0 & 0 \end{pmatrix}.$$

So, in one scattering polarization, say $X'X'$, all $A_g$ and $B_{1g}$ excitations will be presented in one experimental spectrum. Note that the spectra may contain also weak traces of $B_{2g}$ and $B_{3g}$ symmetry leaking from forbidden polarizations due to the wide aperture of the collecting spectrometer optics.

In Fig. 3 we present Raman spectra at selected temperatures. With decreasing temperature, there are drastic changes. In the following, we will focus on the two systems separately.

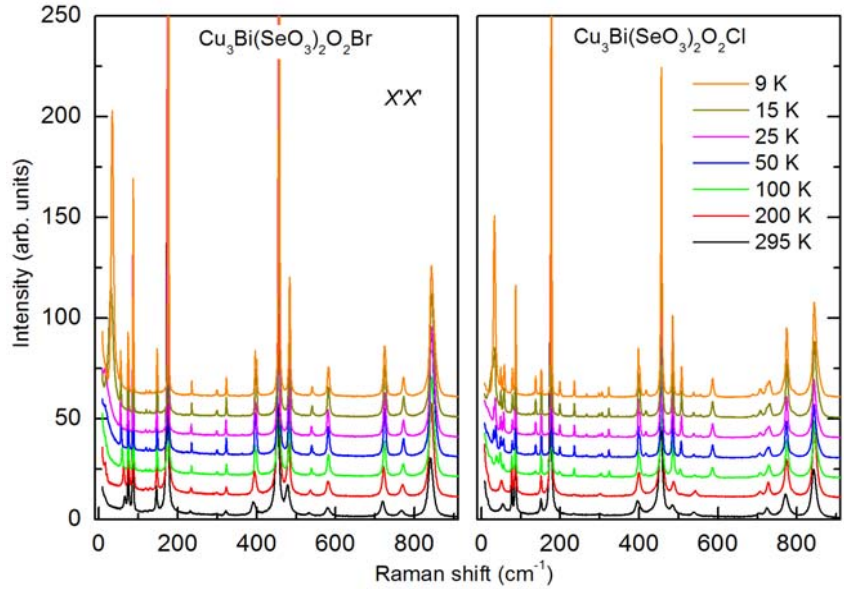

Fig. 3. Temperature dependent $X'X'$ Raman spectra of Cu$_3$Bi(SeO$_3$)$_2$O$_2$Br and Cu$_3$Bi(SeO$_3$)$_2$O$_2$Cl.

**Cu$_3$Bi(SeO$_3$)$_2$O$_2$Br**: In the temperature regime 295 K > $T$ > $T_N$, all phonon lines (except two lines at lowest frequency) show a temperature dependence of the frequency that is expected due to cubic anharmonic interaction. The solid line in Fig. 4(a) is an example of a fit to the $A_g$ phonon frequency using

$$\omega_j(T) = \omega_{0j} - \frac{b_j}{\exp(\hbar\omega_{av3}/k_BT)-1} - \frac{c_j}{\exp(\hbar\omega_{av4}/k_BT)-1} - \frac{c_j}{\left[\exp(\hbar\omega_{av4}/k_BT)-1\right]^2} \quad (1)$$

to include such anharmonic effects.[14, 15, 16] Here, $\omega_{0j}$ is the eigenfrequency of the phonon in the absence of spin-phonon coupling at T = 0 K, $b_j$ and $c_j$ are mode-dependent scaling factors of the anharmonic contributions, and $\omega_{av3}$, $\omega_{av4}$ are averaged frequencies of decayed phonons in three- and four-particle processes (the simplest approximation is $\omega_{av3} \approx \omega_{0j}/2$ and $\omega_{av4} \approx \omega_{0j}/3$). For T ≈ $T_N$ the slope of $\omega(T)$ changes, signalling pronounced spin-phonon interaction.

In Fig. 4(b) the related fit to the phonon linewidth is shown using[15, 16]

$$\Gamma_j(T) = \Gamma_{0j} + \frac{d_j}{\exp(\hbar\omega_{0j}/2k_BT)-1} + \frac{f_j}{\exp(\hbar\omega_{0j}/3k_BT)-1} + \frac{f_j}{\left[\exp(\hbar\omega_{0j}/3k_BT)-1\right]^2}, \quad (2)$$

with $\Gamma_0$, $\omega_0$, $d_j$, and $f_j$ being width, eigenfrequency at T = 0 K, and mode dependent fit parameters for three- and four-particle decay and coalescence processes, respectively. Good agreement with experimental dependence is achieved with $\Gamma_0$ = 2.5, $\omega_0$ = 456, $d$ = 2, and $f$ = 0.5. The integrated intensity of the line is shown in Fig. 4(c). Its abrupt rising below $T_N$ reflects the presence of noticeable spin-phonon interaction in the system.

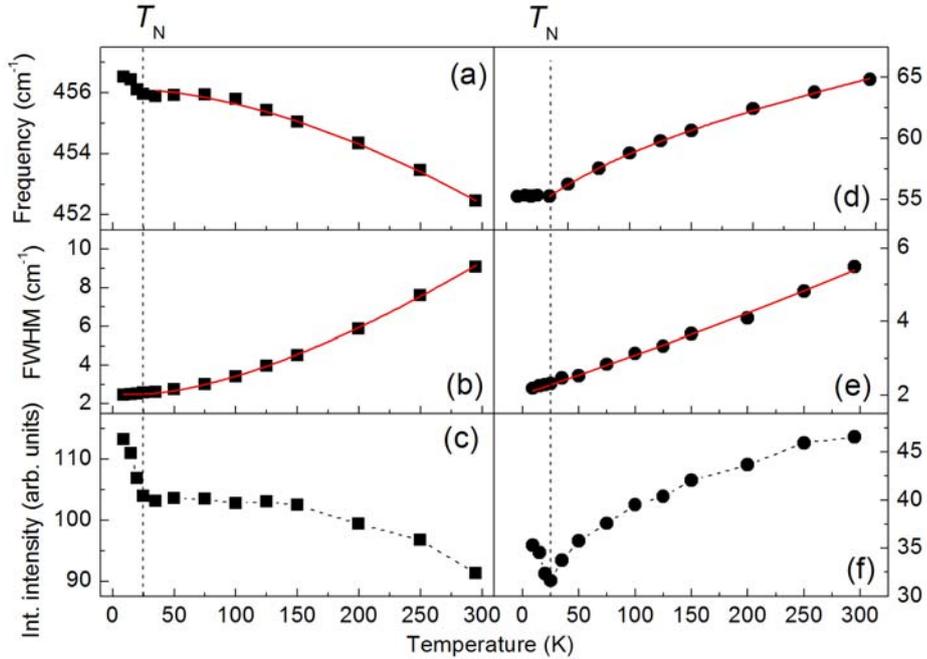

Fig. 4. Parameters of two selected phonon modes in $Cu_3Bi(SeO_3)_2O_2Br$. (a) and (d) Temperature dependence of the frequency. (b) and (e) Linewidth (FWHM) of the modes. (c) and (f) Integrated intensity of the modes. Solid lines in (a-e) present the fit as described in the text.

As mentioned above, the two lowest energy phonon modes at 65 cm$^{-1}$ and 74 cm$^{-1}$ (at RT) are anomalous. The phonon at 74 cm$^{-1}$ slightly softens with decreasing temperature and then hardens for temperatures below $T_N$. The behaviour of the $B_{1g}$ phonon line at 65 cm$^{-1}$ is even more anomalous (see Fig. 4). Its frequency softens by approximately 10 cm$^{-1}$ with cooling from RT to $T_N$. Thereby it follows a square root law, $\omega(T) = \omega_0 + \sqrt{\alpha T}$ (solid line in Fig. 4(d)), where $\omega_0 = 50.18$ cm$^{-1}$ and $\alpha = 0.74$ cm$^{-2}$K$^{-1}$ are the best fitting parameters. For temperatures below $T_N$ the frequency of this phonon line is nearly temperature independent. It is unlikely that such a mode softening could ever result from a structural phase transition. We find a noteworthy similarity of this softening with observations in compounds with rutile structure (see Ref. 17 and references therein). In these systems, lattice thermal contraction disproportionally influences the forces between nearest-neighbour ions in adjacent layers perpendicular to the *c* axis of the crystal.

Also temperature dependence of the phonon intensity is anomalous. It grows for temperatures below $T_N$ (Fig. 4(f)), which is a distinctive feature of spin-phonon coupling. Well above $T_N$ the mode has a symmetric line shape which becomes increasingly asymmetric to higher frequency when the temperature is lowered below $T \approx 75$ K. On the other hand, we did not notice any evidence for a structural phase transition in the temperature range 295 K > $T$ > $T_N$.

Figure 5 zooms into the low-energy part of the Raman spectra. Temperature-dependent quasielastic scattering that extends up to nearly 100 cm$^{-1}$ is evident for T > $T_N$. This quasielastic scattering is attributed to fluctuations of the energy density of the spin system. This mechanism leads to a central peak with Lorentzian lineshape and is inherent to systems with non-negligible spin-phonon coupling, similar to the evidence shown in Fig. 4. The spectral weigh of the quasielastic scattering in $Cu_3Bi(SeO_3)_2O_2Br$ increases upon cooling up to $T_N$ indicating increasing fluctuations of the spin system.

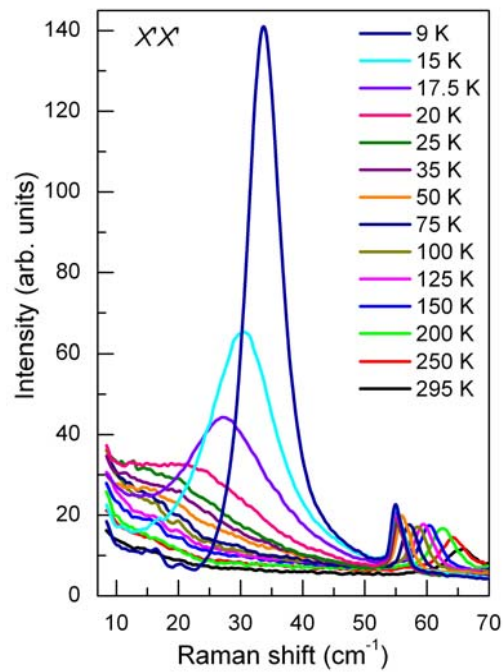

Fig. 5. Low frequency regime of the Raman spectra of $Cu_3Bi(SeO_3)_2O_2Br$.

For temperature below approximately 50 K an additional signal appears on the background of quasielastic scattering which transforms into a mode of distinct, finite-energy at low temperatures. We have tentatively assigned this excitation at 34.5 cm$^{-1}$ to spin degrees of freedom. Note that a magnon excitation with the same energy was observed in an IR study of $Cu_3Bi(SeO_3)_2O_2Cl$.[6] The spectral lineshape of this magnetic mode is asymmetric which we take as evidence of strong coupling with the neighbouring phonon mode.

**Cu$_3$Bi(SeO$_3$)$_2$O$_2$Cl**: For temperatures below $T^* = 120$ K we detect additional phonon modes. In the frequency regime above 80 cm$^{-1}$ and at $T = 100$ K quite intensive lines are located at 81, 198, 200, 416, 504, 555, 586, 688, and 729 cm$^{-1}$. Modes with smaller intensity appear at 120, 137, 267, 299, 324, and 348 cm$^{-1}$. Earlier studies proposed a centrosymmetric-to-noncentrosymmetric, second-order structural phase transition near $T = 115$ K based on the observation of Raman active modes in IR spectra of Cu$_3$Bi(SeO$_3$)$_2$O$_2$Cl.[6] However, no direct evidence for a lowering of crystallographic symmetry for temperatures below 115 K have been found in powder x-ray diffraction.[6] Candidates for low temperature structures are $Pm2_1n$ or $P2_1mn$.

Our Raman observations confirm the existence of a structural phase transition. The observation of a sufficient number of modes (about 9) coinciding in energy with IR bands tends us to accept the loss of an inversion centre in Cu$_3$Bi(SeO$_3$)$_2$O$_2$Cl proposed in Ref. 6. In the case of such a transition $B_{2u}$ and $B_{3u}$ TO modes may appear in the Raman spectra at our experimental conditions. For the proposed low-temperature symmetry $2mm$ they will have polarizations corresponding to the former $B_{1g}$ and $A_g$ phonons, respectively, or vice versa for $m2m$ symmetry. On the other hand former Raman-active phonons of $A_g$ and $B_{1g}$ symmetry must appear in IR data in the appropriate $E\|a$ and $E\|b$ polarizations. A comparison of the joint phonon lines suggests that the more preferable symmetry of the low-temperature phase is $P2_1mn$. In such a case and based on polarization rules of the high-symmetry the correlation scheme of IR/Raman reciprocal intensity leakage of phonon modes is: $A_g \leftrightarrow B_{3u}$, $B_{1g} \leftrightarrow B_{2u}$, $B_{2g} \leftrightarrow B_{1u}$, and $B_{3g} \leftrightarrow A_u$. Table 1 gives a comparison of the modes.

Another argument in favour of a loss of inversion follows from the temperature dependence of the dielectric permittivity of Cu$_3$Bi(SeO$_3$)$_2$O$_2$Cl for $T < T = 50$ K. In the Br compound no such effect exist (Fig. 1(b)). A ferroelectric phase assumes the additional channel of spin-dependent polarizability connected with renormalization of the instant electrical moment by magnetic ordering. Polarizability as the derivative of the moment with respect to an electrical field will have maximum at the temperature with maximum fluctuations. Here, these fluctuations may be provoked by spin fluctuations in the vicinity of $T_N$.

Table I. Phonon frequencies observed in IR from Ref. 6 and in Raman scattering within our study at selected temperatures above and below the structural phase transition. Additional lines (*) appear in the spectra for temperatures below $T^* \approx 120$ K. Modes with an energy coincidence are shown in bold.

| IR from Ref. 6 | | Raman | | | |
|---|---|---|---|---|---|
| 7 K | | 9 K | | 100 K | 295 K |
| $B_{3u}$ | $B_{2u}$ | $A_g$ | $B_{1g}$ | $A_g+B_{1g}$ | $A_g+B_{1g}$ |
| | | 31.1* | | 24.5* | |
| | | | 33.6* | 37.8* | |
| | 36.3 | 38.3* | | 40.8* | |
| | | | 48.6 | 48.1 | 54.8 |
| 52.8 | | 57.8* | 55.6* | 55.9* | |
| 69.9* | 68.3 | | | | |
| | | 78.2 | 78.1 | 77.0 | 77.4 |
| | | 81.4* | | 80.6* | |
| 89.0 | | 86.2 | | 86.4 | 86.3 |
| 101.1* | 99.8* | | | | |
| | 115.2 | | 119.7* | 120.2* | |
| | **128.9*** | | **130.8** | **131.1** | **130.5** |
| | 133.5 | | | | |
| **137.6** | | **137.5*** | | **137.1*** | |
| | | 150.2 | | 151.8 | 151.0 |
| | | 151.6 | | | |
| 161.9 | | | | | |
| 172.3* | | 176.8 | | 176.8 | 175.1 |
| | 185.8 | | 189.2* | 189.0* | |
| 191.6 | | 198.0* | | 197.6* | |
| **202.1** | | **200.4*** | | **200.5*** | |
| | | 235.3 | | 235.4 | 233.0 |
| | 256.9 | | 266.4* | 266.8* | |
| | 276.1* | | | | |
| | **300.3** | | **297.8*** | **298.7*** | |
| | | 305.6 | | 304.4 | 299.0 |
| 320.0* | 313.9 | 322.5* | | 324.1* | |
| 331.1 | | | | | |
| | | 347.5* | | 347.5* | |
| | | | 397.5 | 397.6 | 396.3 |
| | | 400.8 | | 401.6 | |
| | | 417.2* | | 416.2* | |
| 422.9 | | | | | |
| | 456.3 | 455.9 | | 456.9 | 455.0 |
| 470.2 | | | | | |
| | **484.7*** | | **484.7** | **486.2** | **485** |
| | 507.0 | | 507.1* | 504.3* | |
| 542.4* | 542.3 | 537.5 | | 540.8 | 538.6 |
| **557.3** | | **555.5*** | | **552.1*** | |
| | 571.4* | | | | |
| 587.2* | | 586.0* | | 585.8* | |
| **688.2** | | **689.0*** | | **688.0*** | |
| 703.8* | | | | | |

| | 716.1* | | | | |
| --- | --- | --- | --- | --- | --- |
| | | | 723.7 | 724.0 | 724.7 |
| | **730.0** | | 730.2* | 729.0* | |
| 737.0* | | | | | |
| | | 774.2 | | 774.2 | 771.2 |
| 825.0 | 811.3 | | | | |
| | | 844.3 | | 844.8 | 842.4 |

In the following we will turn to the temperature dependence of the phonon modes. Frequency, linewidth and intensity data of two phonons are illustrated in Fig. 6(a-c). The phonon at 455 cm$^{-1}$ with $A_g$ symmetry shows good agreement of its frequency with Eq. 1 (Fig. 6(a)) for temperatures above $T^*$. For temperatures below $T^*$ the frequency first decreases and then increases approaching the magnetic ordering temperature. In general, the observed behaviour in the temperature regime $T_N < T < T^*$ is consistent with the temperature behaviour of the $a$ and $b$ lattice parameters reported in Ref. 6. At $T_N$ we observed a jump of the frequency which is associated with the onset of magnetic ordering. The line width (Fig. 6(b)) is well described by Eq. 2 with a small deviation when approaching $T_N$. The integrated intensity (Fig. 6(c)) shows features associated with structural and magnetic transitions.

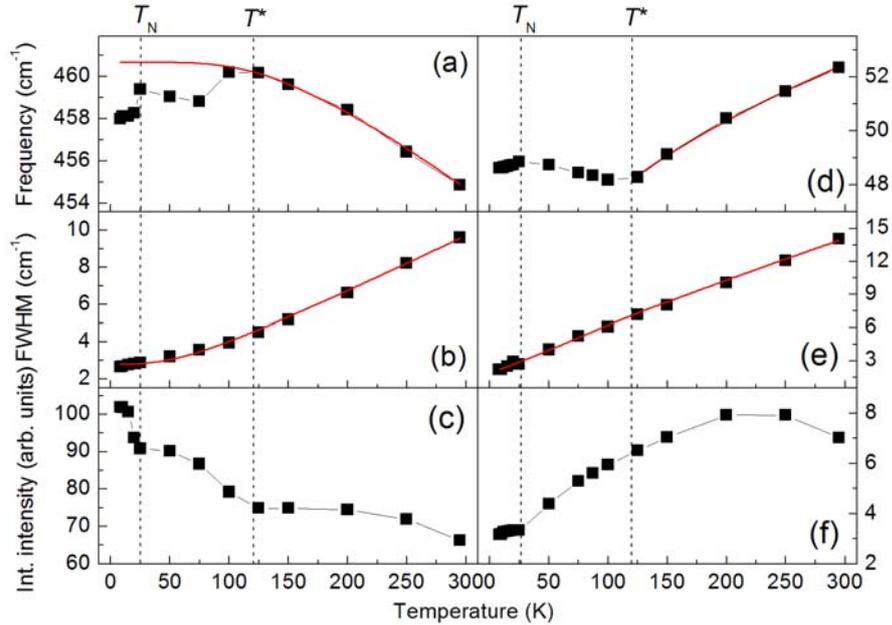

Fig. 6. Parameters of two phonon modes in $Cu_3Bi(SeO_3)_2O_2Cl$. (a) and (d) temperature dependence of the frequency (symbols). (b) and (e) linewidth (FWHM). (c) and (f) integrated intensity. Solid red lines in (a,b,d,e) present fits as described in the text.

In the right panel of Fig. 6 we focus on the data of a low frequency $B_{1g}$ phonon of $Cu_3Bi(SeO_3)_2O_2Cl$. The temperature evolution of its parameters is rather peculiar. At room temperature its frequency is lower by ~8 cm$^{-1}$ compared to the frequency of the corresponding mode in $Cu_3Bi(SeO_3)_2O_2Br$ and its line width is twice larger. With decreasing temperature this mode firstly softens, than hardens below $T^*$, and it is nearly temperature independent below $T_N$ (Fig. 6(d)). This temperature dependence also fits very well to a square root dependence in the temperature regime between RT and $T^*$. The line width (Fig. 6(e)) fits well to Eq. 2 with a small deviation at lower temperatures. The integrated intensity (Fig. 6(f)) mainly responds to $T_N$.

In Fig. 7 zooms into the low-energy regime of Raman spectra in $X'X'$ scattering geometry. There exist an intriguing temperature evolution. For clarity, we present three separate temperature ranges: RT to $T^*$ (Fig. 7(a)), $T^*$ to $T_N$ (Fig. 7(b)), and $T_N$ to 8.5 K (Fig. 7(c)). With cooling, new modes at 38, and 57 cm$^{-1}$ emerge when crossing $T^*$ (Fig. 7(b)). The first mode softens by 3 cm$^{-1}$ while the second one hardens by 0.5 cm$^{-1}$ with decreasing temperatures down to $T_N$. A wide, 30 cm$^{-1}$ width, band which is intensive in $XX$ scattering geometry and has a maximum at 24 cm$^{-1}$ also emerges for $T = 100$ K. This band narrows and shifts to 31.5 cm$^{-1}$ when cooling to $T_N$. We attribute this band to a structural soft mode whose frequency decreases substantially as the transition temperature is approached from below. Fig. 8(c) shows a square frequency temperature dependence in the vicinity of the structural phase transition. We point to the fact also that with a further decrease of temperature the soft mode frequency in $Cu_3Bi(SeO_3)_2O_2Cl$ essentially deviates from a usual law. It is impossible to fit it in the whole temperature range $T^*$-$T_N$ (Fig. 8) with any reasonable assumption of a power law due to interaction with another excitations. Pure $XX$ and $YX$ spectra show that below the structural phase transition the mode at 57 cm$^{-1}$ is presented in two components of the scattering tensor. Their frequencies become clearly different only far below $T_N$. The most intensive mode with $YX$ polarization at 38 cm$^{-1}$ also has a less intensive pair in $XX$ polarization, which demonstrate no temperature dependence, and which becomes visible during the softening of mode $YX$ polarization. So, for temperatures below the structural phase transition 2 modes of $B_{1g}$ and 3 of $A_g$ polarization appear in the frequency regime below 80 cm$^{-1}$.

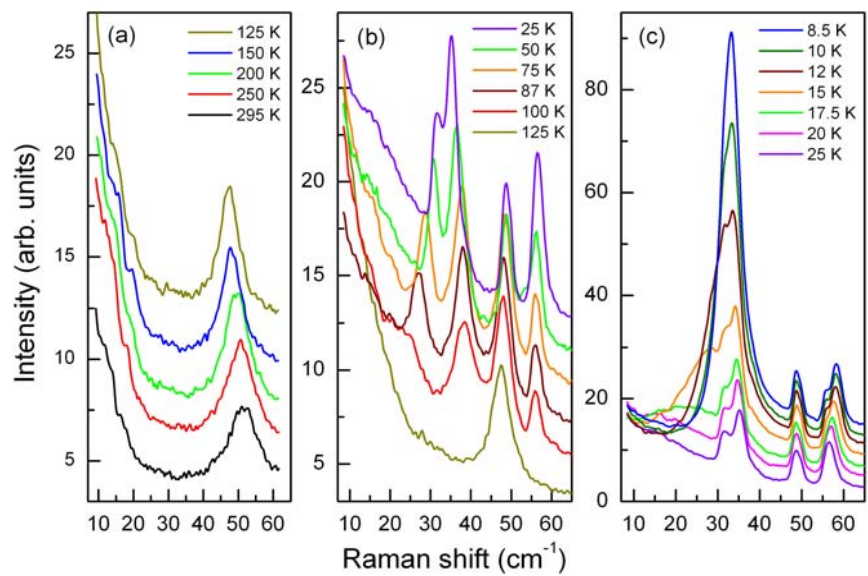

Fig. 7. Low frequency part of $X'X'$ Raman spectra of $Cu_3Bi(SeO_3)_2O_2Cl$. Three temperature ranges are shown: (a) RT to $T^*$, (b) $T^*$ to $T_N$, and (c) $T_N$ to 8.5 K.

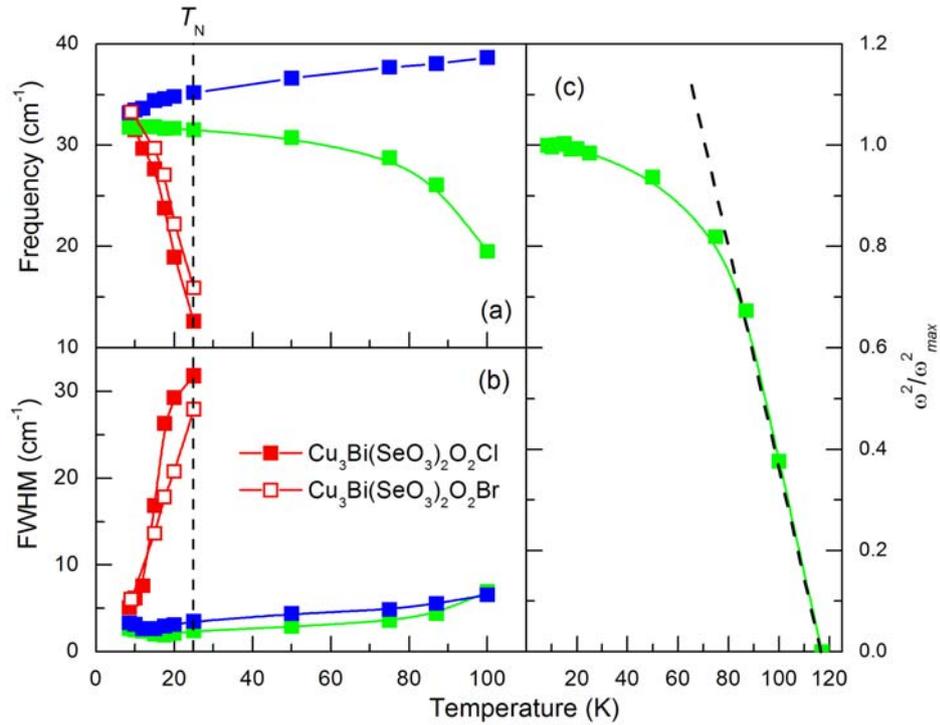

Fig. 8. Temperature dependence of (a) frequency and (b) width of three low frequency modes of $Cu_3Bi(SeO_3)_2O_2Cl$. (c) Soft mode behaviour below $T^*$ with a square frequency dependence of the 31 cm$^{-1}$ mode in $Cu_3Bi(SeO_3)_2O_2Cl$. The data of $Cu_3Bi(SeO_3)_2O_2Br$ are shown for comparison by open symbols.

Now we focus on the thermal behaviour of quasielastic scattering. In $Cu_3Bi(SeO_3)_2O_2Br$ the quasielastic signal growths in intensity with cooling to the magnetic ordering temperature and then decreases (Fig. 9). In contrast, in $Cu_3Bi(SeO_3)_2O_2Cl$ the quasielastic response grows in intensity at $T^*$. The last behaviour, characteristic for the "central" mode behaviour (see Ref. 18), confirms once again the existence of a structural phase transition in $Cu_3Bi(SeO_3)_2O_2Cl$ at $T^* \approx 120$ K.

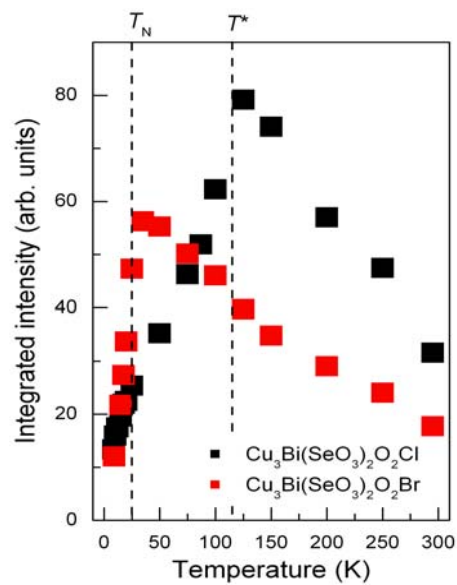

Fig. 9. Temperature dependence of the quasielastic response in $Cu_3Bi(SeO_3)_2O_2Cl$ and $Cu_3Bi(SeO_3)_2O_2Br$. Characteristic temperatures of the compounds are given by dashed lines.

Reconsidering the experimental evidence for a second order structural phase transition in $Cu_3Bi(SeO_3)_2O_2Cl$ at $T^* \approx 120$ K we notice that a specific soft mode behavior in the high symmetry phase *Pmmn* ($T > T^*$) pointing to its *u*-type symmetry has not been observed. However, we surely detected a fully symmetric soft mode of the low symmetry phase (Fig. 7 and 8). Therefore the low-temperature phase belongs to the $P2_1mn$ space group in which *A* type Raman active phonons are also FIR active. Such a space group symmetry also implies that $Cu_3Bi(SeO_3)_2O_2Cl$ is **multiferroic** for $T < T^* \approx 120$ K with a static electric dipole moment directed along the crystallographic *a* axis.

At low temperatures yet another band with lowest-frequency appears in the Raman spectra of $Cu_3Bi(SeO_3)_2O_2Cl$ (Fig. 7(c)). It narrows, hardens and merges with two neighbouring lines into a single asymmetric line in *X'X'* spectra at lowest temperatures. This excitation is suggested to be of magnetic origin. Also IR data has revealed a magnetic excitation at 33.1 cm$^{-1}$ for $T < T_N$.[6] Its magnetic origin is confirmed by its polarization and magnetic field dependence.

## IV. SYMMETRY CONSIDERATIONS

In the following we will discuss magnetic neutron and Raman scattering in $Cu_3Bi(SeO_3)_2O_2X$, X= Br and Cl. Details of long range magnetic order of $Cu_3Bi(SeO_3)_2O_2Br$ have been defined in neutron powder diffraction experiments.[4] The magnetic unit cell is formed by crystallographic unit cell of the *Pmmn* space group which is doubled along *c*-axis in accordance with magnetic propagation vector $\boldsymbol{k} = (0, 0, ½)$. Thus eight Cu1 and four Cu2 ions form twelve branches of usual transversal linear spin waves. From these modes nine branches of exchange magnons have energies which are defined by exchange interactions.[1] They can appear in low-frequency Raman scattering due to one magnon scattering. Their intensities can be very large because of the highly noncollinear structure.

In spite of the structural phase transition in $Cu_3Bi(SeO_3)_2O_2Cl$ the following symmetry considerations of magnetic scattering remain relevant even in the low symmetry phase. This consideration is possible due to the second order structural phase transition and therefore *Pmmn* → *P2₁mn* group-subgroup relation. This is an advantage of the irreducible representation analysis.

To analyze the symmetry of the magnetic degrees of freedom we introduce linear combinations of Fourier transformed sublattices spin $\boldsymbol{s}_\alpha(\boldsymbol{k})$ for the Cu1 and $\boldsymbol{s}_\beta(\boldsymbol{k})$ for the Cu2 sites where $\alpha = 1 \div 4$ and $\beta = 5, 6$.

$$\begin{aligned}
\boldsymbol{F}(\boldsymbol{k}) &= 1/4\left(\boldsymbol{s}_1(\boldsymbol{k}) + \boldsymbol{s}_2(\boldsymbol{k}) + \boldsymbol{s}_3(\boldsymbol{k}) + \boldsymbol{s}_4(\boldsymbol{k})\right); \quad \boldsymbol{m}(\boldsymbol{k}) = 1/2\left(\boldsymbol{s}_5(\boldsymbol{k}) + \boldsymbol{s}_6(\boldsymbol{k})\right); \\
\boldsymbol{L}_1(\boldsymbol{k}) &= 1/4\left(\boldsymbol{s}_1(\boldsymbol{k}) + \boldsymbol{s}_2(\boldsymbol{k}) - \boldsymbol{s}_3(\boldsymbol{k}) - \boldsymbol{s}_4(\boldsymbol{k})\right); \quad \boldsymbol{l}(\boldsymbol{k}) = 1/2\left(\boldsymbol{s}_5(\boldsymbol{k}) - \boldsymbol{s}_6(\boldsymbol{k})\right); \\
\boldsymbol{L}_2(\boldsymbol{k}) &= 1/4\left(\boldsymbol{s}_1(\boldsymbol{k}) - \boldsymbol{s}_2(\boldsymbol{k}) + \boldsymbol{s}_3(\boldsymbol{k}) - \boldsymbol{s}_4(\boldsymbol{k})\right); \\
\boldsymbol{L}_3(\boldsymbol{k}) &= 1/4\left(\boldsymbol{s}_1(\boldsymbol{k}) - \boldsymbol{s}_2(\boldsymbol{k}) - \boldsymbol{s}_3(\boldsymbol{k}) + \boldsymbol{s}_4(\boldsymbol{k})\right);
\end{aligned} \quad (1)$$

We use the following numbering of the copper ions positions: for Cu1 {1 – (0, 0, 0); 2 – (0.5, 0.5, 0); 3 – (0, 0.5, 0); 4 – (0.5, 0, 0)} and for Cu2 {5 – (0.25, 0.25, 0.79); 6 – (0.75, 0.75, 0.21) }. In a case of a second-order magnetic phase transition, the possible magnetic structures can be classified by the irreducible representations (IRP) of the symmetry group of the crystal in the paramagnetic phase. The basis functions of IRP are formed from Cartesian components of vectors (1). To account for the doubling of the primitive cell in the ordered phase we should consider magnetic degree of freedom of the paramagnetic phase at two wave vectors $\boldsymbol{k}_{(+)} = (0, 0, 0)$ and $\boldsymbol{k}_{(-)} = (0, 0, 1/2)$. Fortunately the set of IRP coincides with each other for small group of $\boldsymbol{k}_{(+)}$ and $\boldsymbol{k}_{(-)}$ wave vectors. Below we will use the notation (+) or (−) to denote IRP and magnetic basis functions with respective $\boldsymbol{k}_{(+)}$ and $\boldsymbol{k}_{(-)}$ wave vectors. The symmetry classification is shown in Table II.

Table II. Magnetic basis functions in $Cu_3Bi(SeO_3)_2O_2X$ ( X = Br, Cl). The calculation of the IRP is based on the procedure described in Ref. 19.

|  |  | Cu1 | Cu2 |  |
|---|---|---|---|---|
| IRP | | $k_{(+)}, k_{(-)}$ | $k_{(+)} = (0, 0, 0)$ | $k_{(-)} = (0, 0, 1/2)$ |
| $\Gamma_1^{(+)}, A_g$ | $\Gamma_1^{(-)}$ | $L_{3x}^{(\pm)}, L_{2y}^{(\pm)}, L_{1z}^{(\pm)}$ | - | - |
| $\Gamma_2^{(+)}, A_u$ | $\Gamma_2^{(-)}$ | - | $l_z^{(+)}$ | $m_z^{(-)}$ |
| $\Gamma_3^{(+)}, B_{1g}$ | $\Gamma_3^{(-)}$ | $L_{2x}^{(\pm)}, L_{3y}^{(\pm)}, F_z^{(\pm)}$ | $m_z^{(+)}$ | $l_z^{(-)}$ |
| $\Gamma_4^{(+)}, B_{1u}$ | $\Gamma_4^{(-)}$ | - | - | - |
| $\Gamma_5^{(+)}, B_{2g}$ | $\Gamma_5^{(-)}$ | $L_{1x}^{(\pm)}, F_y^{(\pm)}, L_{3z}^{(\pm)}$ | $m_y^{(+)}$ | $l_x^{(-)}$ |
| $\Gamma_6^{(+)}, B_{2u}$ | $\Gamma_6^{(-)}$ | - | $l_x^{(+)}$ | $m_y^{(-)}$ |
| $\Gamma_7^{(+)}, B_{3g}$ | $\Gamma_7^{(-)}$ | $F_x^{(\pm)}, L_{1y}^{(\pm)}, L_{2z}^{(\pm)}$ | $m_x^{(+)}$ | $l_y^{(-)}$ |
| $\Gamma_8^{(+)}, B_{3u}$ | $\Gamma_8^{(-)}$ | - | $l_y^{(+)}$ | $m_x^{(-)}$ |

Neutron diffraction studies reveal that magnetic long range order in $Cu_3Bi(SeO_3)_2O_2Br$ has a symmetry which is described by a $\Gamma_3^{(-)}$ irreducible representation[4]. This is in accordance with concept of "one irreducible representation" which should be realized for a second order phase transition. It means that emergent nonzero magnetic order parameters are $\bar{L}_{2x}^{(-)}$, $\bar{L}_{3y}^{(-)}$, $\bar{F}_z^{(-)}$, $\bar{l}_z^{(-)}$. The respective magnetic space group is $P_{2c}mmn$ (N #59.8.485 in Opechowski-Guccione setting). From similar static magnetic properties it is supposed that $Cu_3Bi(SeO_3)_2O_2Cl$ has the same magnetic symmetry as $Cu_3Y(SeO_3)_2O_2Cl$.[20]

In accordance with this analysis symmetry allowed Dzyaloshinsky-Moriya interactions (DMI) must be present both along Cu1-Cu1 and Cu1-Cu2 bonds. As usual in rhombic antiferromagnets with only one dominate AFM exchange only one of order parameters is important whereas others are smaller with a magnitude that is given by the ratio of DMI to the dominate exchange. In francisites the $\bar{L}_{3y}^{(-)}$, $\bar{F}_z^{(-)}$ and $\bar{l}_z^{(-)}$ order parameters have the same order of magnitude due to competing FM NN and frustrating AFM NNN exchange interactions and DMI plays a minor role in the formation of large noncollinearity.

This also means that in quantum magnetic fluctuations of all $L_3^{(-)}$, $F^{(-)}$ and $l^{(-)}$ order parameters will be present on an equal footing.

Exchange interactions have higher symmetry because the scalar product of magnetic moments depends just on their mutual angles. The symmetry of exchange multiplets are listed in Table III. The most unusual result shown here is the presence of the exchange multiplet $l^{(+)}$ which has rotational symmetry of $P_z$ – electric dipole moment directed along the $c$ axis. This is possible as the Cu2 ion site is not inversion symmetric despite the centre of inversion of the *Pmmn* space group. Similarly, Table II contains magnetic basis functions of Cu2 ions which transform according to the $\Gamma_6^{(+)}$ and $\Gamma_8^{(+)}$ IRP's e.g. they have rotational symmetry of the $P_y$ and $P_x$ components of electric dipole moments respectively. We will see that these circumstances have drastic impact on the following spectroscopy features.

Table III. Symmetry of exchange multiplets.

|  | Cu1 | Cu2 | |
|---|---|---|---|
| IRP | $k_{(+)}, k_{(-)}$ | $k_{(+)} = (0, 0, 0)$ | $k_{(-)} = (0, 0, 1/2)$ |
| $\Gamma_1^{(\pm)}$ | $F^{(\pm)}$ | $m^{(+)}$ | $l^{(-)}$ |
| $\Gamma_2^{(\pm)}$ | - | - | - |
| $\Gamma_3^{(\pm)}$ | $L_1^{(\pm)}$ | - | - |
| $\Gamma_4^{(\pm)}$ | - | $l^{(+)}$ | $m^{(-)}$ |
| $\Gamma_5^{(\pm)}$ | $L_2^{(\pm)}$ | - | - |
| $\Gamma_6^{(\pm)}$ | - | - | - |
| $\Gamma_7^{(\pm)}$ | $L_3^{(\pm)}$ | - | - |
| $\Gamma_8^{(\pm)}$ | - | - | - |

Using Table III one can consider the strongest exchange contribution to magnetic Raman scattering in francisites. Respective invariants are shown in Table IV for different Raman symmetry channels.

Table IV. Exchange magnetic contribution into Raman tensors with origin from different Cu sites

| IRP | Cu1-Cu1 | Cu1-Cu2 |
|---|---|---|
| $A_g$ | $\sum_{v=0,1,2,3} \Pi_{ii}^{(v)}\left(k_{(\pm)}\right)\left(L_v^{(\pm)}\right)^2$ | $\pi_{ii}^{(+)}\left(k_{(+)}\right)\left(m^{(+)} \cdot F^{(+)}\right) + \pi_{ii}^{(-)}\left(k_{(-)}\right)\left(l^{(-)} \cdot F^{(-)}\right)$ |
| $B_{1g}$ | $\Pi_{XY}\left(k_{(\pm)}\right)\left(L_2^{(\pm)} \cdot L_3^{(\pm)}\right)$ | - |
| $B_{2g}$ | $\Pi_{XZ}\left(k_{(\pm)}\right)\left(L_3^{(\pm)} \cdot L_1^{(\pm)}\right)$ | - |
| $B_{3g}$ | $\Pi_{YZ}\left(k_{(\pm)}\right)\left(L_2^{(\pm)} \cdot L_1^{(\pm)}\right)$ | - |

Here the constants $\Pi_{ij}^{(\pm)}$ and $\pi_{ij}^{(\pm)}$ are constructed from the polarizability of the pair of nearest neighbors Cu1-Cu1 ions and Cu1-Cu2 ions, respectively. Due to large distances between the Cu2 ions the contribution from the Cu2-Cu2 pairs should be neglected. Cartesian indices *ij* denote polarizations of the incident and scattered light, respectively. As follows from the symmetry analysis two-magnon Raman scattering in francisites must exist in all polarization channels. Note that spin-dependent parts for all channels are different and therefore two-magnon scattering should demonstrate polarization dependent anisotropic behavior.

However, also very intensive one magnon Raman scattering caused by the exchange mechanisms shown in Table IV is possible in francisites. This unique situation is based on the large noncollinearity of the magnetic structure with large angle $\theta$ between the magnetic moments of Cu1 and Cu2 sublattices. In $Cu_3Bi(SeO_3)_2O_2Br$ $\theta \approx 51.6°$, where the angle value is extracted from the ratio $\bar{L}_{3y}^{(-)}/\bar{F}_z^{(-)} = tg\theta$ for Cu1 magnetic moments, while Cu2 magnetic moments are directed along the *z*-axis.[4] The huge canting angle depends mainly on the relation between Cu1-Cu2 ferromagnetic exchange and frustrating NNN Cu1-Cu1 AFM exchange.[1] Such kind of canting, which has nonrelativistic origin, supports the scattering on exchange magnons in the multisublattices exchange noncollinear magnets.[21] The one magnon contribution from exchange mechanisms appears under replacement one of the basis functions in the invariants shown in Table IV by nonzero magnetic order parameters. Respective selection rules can be directly derived from this condition if the symmetry of spin waves at *k* = 0 is known. This mechanism does not exist in collinear magnets.

To analyze the symmetry of the linear spin wave spectra one should investigate the content of the so called spin-wave representation. For one magnon scattering the spin waves solely at the Γ point of the magnetic Brillouin zone have to be considered. Note, that in the magnetic phase the paramagnetic Brillouin zone is folded along z-axis, so that $k_{(-)} = (0, 0, ½)$ point is the Γ point in the magnetic Brillouin zone. In Table V we classify magnons at the magnetic Γ point by IRP of paramagnetic phase. This classification implies that small spin deviations can be described by magnetic basis functions (1). Then the equations of motion for spin operators in form of the magnetic basis function (1) and following linearization of these equations makes it possible to indicate which small deviations of (1) take part in the given branch of spin waves.[22] The actual magnetic symmetry is accounted for by the concrete IRP of the magnetic ground state under linearization-decoupling procedure.

Table V. Symmetry, type and spectroscopic activity of linear spin waves in the center of the magnetic Brillouin zone in $Cu_3Bi(SeO_3)_2O_2X$ (X=Cl, Br). AM and EM denote acoustic- and exchange-type magnons, respectively.

| IRP | Type and sublattice participation | FIR active (polarization) | Raman active (polarization) |
|---|---|---|---|
| $\Gamma_1^{(-)} \times \Gamma_3^{(+)}$ | 1AM+1EM (Cu1) | $m\|z$ | XY cross polarization, (exchange and Faraday mechanisms) |
| $\Gamma_1^{(+)} \times \Gamma_3^{(-)}$ | 2EM (Cu1) | - | Parallel polarization |
| $\Gamma_5^{(+)} \times \Gamma_7^{(-)}$ | 1AM+2EM (Cu1+Cu2) | $m\|y$ | XZ cross polarization (due to Faraday mechanism) |
| $\Gamma_5^{(-)} \times \Gamma_7^{(+)}$ | 1AM+2EM (Cu1+Cu2) | $m\|x$ | YZ cross polarization (due to Faraday mechanism) |
| $\Gamma_6^{(-)} \times \Gamma_8^{(+)}$ | 1EM(Cu2) | $p\|x$ | - |
| $\Gamma_6^{(+)} \times \Gamma_8^{(-)}$ | 1EM(Cu2) | $p\|y$ | - |

As follows from Table V there are no silent magnon modes in $Cu_3Bi(SeO_3)_2O_2X$. All of them should be seen in one magnon Raman or FIR studies. Surprising, however two modes are pure electromagnons because they can be excited just by an electric field.

The following analysis shows that two modes which should be seen in parallel Raman polarization have frequencies 13.9 cm$^{-1}$ and 54 cm$^{-1}$. To obtain these values we use a procedure of quantization[21] and data for exchange integrals and magnetization values from Ref. 4. One can show that the scattering intensity of the low lying mode contains an accidentally small parameter which is connected with a ratio of exchange integrals. Meanwhile the scattering intensities on the higher mode include a factor $\left(\overline{L}_{2x}^{(-)}\right)^2$ which has relativistic origin (some ratio of Dzyaloshinskii-Moriya interaction to some combination of exchange integrals.[1] Furthermore, as shown by elastic neutron scattering[4] for $Cu_3Bi(SeO_3)_2O_2Br$, $\overline{L}_{2x}^{(-)} = 0$. Thus the one magnon scattering on the high frequencies spin waves cannot explain the observed magnetic features in our Raman spectra.

The origin of the enormous magnetic Raman intensity can be explained by a Fano resonance of one of the low energy phonons with magnetic fluctuations that develop around the Neel temperature and become gapped at low temperatures. These quantum fluctuations arise as a reminiscence of the manifold of degenerated classical ground states on the AFM kagome lattice which should lead to the onset of an "order by disorder" phenomenon.[23] Specifically in the francisites with FM NN coupling and frustrating AFM NNN coupling, its realization in the isotropic limit is seen due to presence of a line of zero-energy modes along the *b*-axis in momentum space.[1] However, as shown in Ref. 1 the global degeneracy of the isotropic Heisenberg $S=\frac{1}{2}$ model can be removed and zero-energy modes become gapped under accounting of the Dzyaloshinsky-Moriya interactions. Below we analyze the so called Fano terms which allow some mixture of magnetic fluctuations (or two magnon continua) with phonons in Raman scattering.

Table VI. Fano terms of exchange origin for spin lattice interference contribution into Raman tensors. Here the constants $T^\nu$ are the normal coordinate, $Q^\nu$ are derivatives of the pair polarizabilities $\Pi$ shown in Table IV.

| Phonon type involved | Cu1-Cu2 |
|---|---|
| $A_g$ | $\sum_\nu T_{ii}^{\nu A_g}\left(k_{(-)}\right)\left(F^{(-)} \cdot l^{(-)}\right) Q_{A_g}^\nu (0)$ |
| $B_{1u}$ ($P_z$) | $\sum_\nu T_{ii}^{\nu B_{1u}}\left(k_{(-)}\right)\left(F^{(-)} \cdot m^{(-)}\right) Q_{B_{1u}}^\nu (0)$ |
| $B_{2u}$ ($P_y$) | $\sum_\nu T_{ii}^{\nu B_{2u}}\left(k_{(-)}\right)\left(L_3^{(-)} \cdot m^{(-)}\right) Q_{B_{2u}}^\nu (0)$ |
| $B_{3u}$ ($P_x$) | $\sum_\nu T_{ii}^{\nu B_{3u}}\left(k_{(-)}\right)\left(L_2^{(-)} \cdot m^{(-)}\right) Q_{B_{3u}}^\nu (0)$ |

The involvement of polar phonons into the Fano process is the most unusual result, see Table VI for Fano terms. It means that Raman scattering from a crystal with inversion symmetry allows the observation of forbidden polar phonons with a Fano resonance to magnetic quantum fluctuations. Notice that in francisites all of these possibilities are connected with the special site symmetry of the Cu2 ions which does not include an inversion.

There are a number of polar phonons in francisites with energies that enable a crossing with gapped magnetic fluctuations.[6, 5] In $Cu_3Bi(SeO_3)_2O_2Cl$ the lowest energy mode corresponds to a $B_{2u}$ TO-phonon with $\Omega_F = 36.3$ cm$^{-1}$ and a huge oscillator strength $S = 26.2$ (e.g. huge dipole moment). Therefore there exist an enormous *LO-TO* splitting that spreads up to 55.6 cm$^{-1}$.[6] This phonon has a large energy overlap with the magnetic fluctuation spectrum and provides an immense Fano line width (Fig. 7 (b)) covering several near phonon lines. Its enormously large dipole moment is suggested to lead to the giant intensity of the Fano resonance. The latter intensity starts to rise when the maximum of the spectral weight of the magnetic fluctuations overlaps with the phonon line. We suppose that a similar scenario of Fano resonance develops in the isostructural $Cu_3Bi(SeO_3)_2O_2Br$ with similar frequencies of the low lying phonons (Fig. 5).[5] We emphasize that such a coupling of polar phonons to magnetic fluctuations is rather special. It means that the joint fluctuations of the Cu1 AFM order parameter $L_3^{(-)}$ and the Cu2 order parameter $m^{(-)}$ carry electric dipole moments.

Below we will discuss an additional possible source of a pronounced magnetic Raman signal in the frequency range of around 30-40 cm$^{-1}$. This is again connected with crucial role of the $B_{2u}$ TO-phonon. This phonon participates in a strong modulation of both $J_F$ NN and $J_{AF}$ NNN exchanges due to large dipole moment and therefore large out-of-phase displacements of positive and negative ions which take part in the exchange bond. In particular, it leads to dimer-like modulations (in spite of FM exchange) of the distance between Cu1 ions along the $a$ axis. As shown in Ref. 1 the magnitude of sublattices magnetic moments depends on the ratio of $J_F/J_{AF}$. Therefore the $B_{2u}$ phonon oscillation yields a variation of sublattices moments with the same $\Omega_F$ frequency through a modulation of exchange interactions. In the Raman Stokes process it leads to an additional channel of scattering, namely to a one magnon scattering process on a longitudinal magnon. The respective signal has to be observed in parallel polarization with high enough intensity as it is provided by exchange scattering $\pi_{ii}^{(-)}\left(k_{(-)}\right)\left(l^{(-)} \cdot F^{(-)}\right)$. Here, to obtain the Raman tensor for one magnon scattering on longitudinal magnons, one of the basis set of spins (1) must be replaced by the mean-field value.[24] It is remarkable, that here the Raman intensity originates from $l^{(-)}$ and $F^{(-)}$ moments which appear due to the magnetic BZ folding. If the translational symmetry is restored by a metamagnetic transition in an external field $H\|c$ simultaneous with the disappearance of such moments also low-frequency magnetic peaks should disappear.

Finally we consider the unusual appearance of an electric dipole moment (e.g. electro magnons) in a crystal with inversion symmetry. Acting in the spirit of Moriya-Tanabe-Sugano approach[25] one can show that the electric dipole moments are allowed for $\Gamma_6^{(-)} \times \Gamma_8^{(+)}$ and $\Gamma_6^{(+)} \times \Gamma_8^{(-)}$ type of magnons even in exchange approximation, see Table VII.

Table VII Electric dipole moment of exchange origin for one- and two-magnon absorption.

| IRP | Cu1×Cu2 |
|---|---|
| $B_{1u}(P_z)$ | $\pi_z\left(k_{(-)}\right)\left(F^{(-)} \cdot m^{(-)}\right)$ |
| $B_{2u}(P_y)$ | $\pi_y\left(k_{(-)}\right)\left(L_3^{(-)} \cdot m^{(-)}\right)$ |
| $B_{3u}(P_x)$ | $\pi_x\left(k_{(-)}\right)\left(L_2^{(-)} \cdot m^{(-)}\right)$ |

Respective operator of the electric dipole moment for one magnon absorption comes from the equations of Table VII by the replacement one of $\boldsymbol{F}^{(-)}$, $\boldsymbol{L}_2^{(-)}$, $\boldsymbol{L}_3^{(-)}$ sets of spins into mean-field value. Notice, that longitudinal magnons can be excited by electric field with $z$-polarization.

## V. CONCLUSIONS

In summary, $Cu_3Bi(SeO_3)_2O_2X$ gives us the opportunity studying lattice and magnetic excitations of a pseudo-kagome systems with and without inversion symmetry. The loss of inversion due to a second order structural phase transition ($T^* \approx 120$ K) in $Cu_3Bi(SeO_3)_2O_2Cl$ is based on: (i) specific heat and dielectric permittivity, (ii) soft mode in Raman of the low-temperature phase, and (iii) reciprocal emergence of phonon lines in Raman and IR spectra. The latter analysis leads to the $P2_1mn$ low-temperature phase. Intensive low-frequency magnetic modes are observed in both $Cu_3Bi(SeO_3)_2O_2Cl$ and $Cu_3Bi(SeO_3)_2O_2Br$. There exist a rather unusual relation of this magnetic signal to a polar phonon. Such a connection points to an electro-dipole character of magnetic fluctuations which exist in a crystal with inversion ($Cu_3Bi(SeO_3)_2O_2Br$), as well as without inversion ($Cu_3Bi(SeO_3)_2O_2Cl$). To elucidate the role of symmetry further we propose experiments in magnetic fields using the higher symmetry of the spin-flop (ferromagnetic) phase. The low energy modes find their explanation in scattering on longitudinal magnons. Such a mode has previously only been observed in the quantum spin system $Cu_2Te_2O_5Br_2$ that is in proximity to a quantum critical point.[26] Our study demonstrates a rather complete picture of magnetic Raman scattering based on different mechanisms that proves the richness of physics of this type of pseudo-kagome spin 1/2 systems.

**Acknowledgements**

This work was supported by the Russian-Ukrainian RFBR-NASU project 78-02-14, in part from the Ministry of Education and Science of the Russian Federation in the framework of Increase Competitiveness Program of NUST «MISiS» № K2-2015-075 and by Act 211 of the Government of Russian Federation, agreement № 02.A03.21.0006. This work was also supported by the NTH-School "Contacts in Nanosystems: Interactions, Control and Quantum Dynamics", the Braunschweig International Graduate School of Metrology (IGSM), and DFG-RTG 1952/1, Metrology for Complex Nanosystems.